\documentclass[12pt,preprint,showpacs,tightenlines,preprintnumbers,superscriptaddress,nofootinbib]{revtex4}
\usepackage{epsf}
\usepackage{amsmath}
\usepackage{mathrsfs}
\usepackage{xcolor}
\usepackage{cancel}

\def\beq{\begin{equation}}
\def\eeq{\end{equation}}
\def\bea{\begin{eqnarray}}
\def\eea{\end{eqnarray}}
\def\beqa{\begin{equation}\begin{array}{l}}
\def\eeqa{\end{array}\end{equation}}
\def\eqlab#1{\label{eq:#1}}
\def\figlab#1{\label{fig:#1}}
\def\tablab#1{\label{tab:#1}}
\def\seclab#1{\label{sec:#1}}

\def\Eqref#1{Eq.~(\ref{eq:#1})}

\def\Figref#1{Fig.~\ref{fig:#1}}
\def\tabref#1{\ref{tab:#1}}

\def\secref#1{\ref{sec:#1}}
\def\sla#1{#1 \!\!\!\! \slash\,}

\def\slad{\partial \hspace{-2.1mm} \slash}

\def\slaa{a \hspace{-2mm} \slash}

\def\half{\mbox{\small{$\frac{1}{2}$}}}

\def\barr{\left(\begin{array}{c}}
\def\earr{\end{array}\right)}
\def\bmat{\left(\begin{array}{cc}}
\def\emat{\end{array}\right)}
\def\al{\alpha}
\def\be{\beta}
\def\ga{\gamma} 
\def\de{\delta} \def\De{\Delta}\def\vDe{\varDelta}
\def\veps{\varepsilon}  \def\eps{\epsilon}

\def\la{\lambda} \def\La{{\Lambda}}

\def\si{\sigma} 
\def\th{\theta}

\def\pa{\partial}

\def\pa{\partial}

\def\nn{\nonumber}

\def\cO{\mathscr{O}}

\def\lag{{\mathcal L}}

\def\ceps{\mathscr{E}}

\def\3d{3-D}

\def\ol#1{\overline{#1}}

\begin{document}

\title{Predictive powers of chiral perturbation theory in Compton scattering off protons}
\author{Vadim Lensky}
\email{lensky@ect.it}

\affiliation{European Centre for Theoretical Studies in Nuclear Physics and Related Areas (ECT*), \\
Villa Tambosi, Villazzano (Trento),
I-38050 TN, Italy}

\affiliation{Institute for Theoretical and Experimental Physics, 117218 Moscow, Russia\footnote{On leave of absence.}}

\author{Vladimir Pascalutsa}

\affiliation{European Centre for Theoretical Studies in Nuclear Physics and Related Areas (ECT*), \\
Villa Tambosi, Villazzano (Trento),
I-38050 TN, Italy}

\affiliation{Institut f\"ur Kernphysik, Johannes Gutenberg Universit\"at, Mainz D-55099, Germany}

\date{\today}

\begin{abstract}
We study low-energy nucleon Compton scattering  in the 
framework of baryon chiral perturbation theory (B$\chi$PT) with
pion, nucleon, and $\Delta$(1232) degrees of freedom, up to and including the
next-to-next-to-leading
order (NNLO). We include the effects of order $p^2$, $p^3$ and $p^4/\varDelta$,
with $\varDelta\approx 300$ MeV the $\Delta$-resonance excitation energy.
These are all ``predictive" powers in the sense that no unknown low-energy constants enter
until at least one order higher (i.e, $p^4$). Estimating the theoretical uncertainty on the basis
of natural size for $p^4$ effects, we find that uncertainty of such a NNLO result is comparable to
the uncertainty of the present experimental data for low-energy Compton scattering.
We find an excellent agreement with the experimental cross section data up
to at least the pion-production threshold. Nevertheless, for
the proton's magnetic polarizability we obtain a value of $(4.0\pm 0.7)\times 10^{-4}$ fm$^3$,
in significant disagreement with the current PDG value.
Unlike the previous $\chi$PT studies
of Compton scattering, we perform the calculations in
a manifestly Lorentz-covariant fashion, refraining from the
heavy-baryon (HB) expansion.  The difference between the lowest order 
HB$\chi$PT and B$\chi$PT results for polarizabilities is found to be appreciable.
We discuss the chiral behavior
of proton polarizabilities in both HB$\chi$PT and B$\chi$PT with the hope
to confront it with lattice QCD calculations in a near future.
In studying some of the polarized observables, we identify the regime where
their naive low-energy expansion begins to break down, thus addressing
the forthcoming precision measurements at the HIGS facility.
\end{abstract}


\maketitle
\thispagestyle{empty}
\tableofcontents

\section{Introduction}

Compton scattering off nucleons has a long and exciting history, 
see Refs~\cite{Drechsel:2002ar,Schumacher:2005an,Phillips:2009af}
for recent reviews. The 90's witnessed a breakthrough in experimental
techniques which led to a series of precision measurements of Compton scattering~\cite{Schmiedmayer:1991zz,Federspiel:1991yd,Zieger:1992jq,Hal93,MacG95,MAMI01}
with the aim to determine the
 nucleon {\it  polarizabilities}~\cite{Baldin,Holstein:1992xr}. 

Many theoretical approaches have been tried in the description of nucleon polarizabilities
and low-energy Compton scattering. The more prominent examples include dispersion relations \cite{Hearn:1962zz,Pfeil:1974ib,Guiasu:1978ak,Lvov:1980wp,L'vov:1996xd,Drechsel:1999rf,Pasquini:2007hf}, effective-Lagrangian models~\cite{Pascalutsa:1995vx,Scholten:1996mw,Feuster:1998cj,Kondratyuk:2000kq},
constituent quark model~\cite{Capstick:1992tx}, and chiral-soliton type of models~\cite{Chemtob:1987ut,Scoccola:1989px,Scherer:1992jb,Broniowski:1992vj,Scoccola:1995tf}.
There has been as well a significant recent progress 
in approaching the subject from first principles---lattice QCD (lQCD).
The present lQCD
studies are based on the external electromagnetic field method~\cite{Lee:2005dq,Detmold:2006vu},
and even though
the actual results for the nucleon have been obtained only in quenched approximation, 
the pion and kaon polarizabilities have been calculated with dynamical quarks~\cite{Detmold:2009dx}.
The full-lQCD calculations for the nucleon will hopefully be done in a near future.

In this work we exploit another theoretical approach rooted in QCD, namely, chiral perturbation theory ($\chi$PT)~\cite{Pagels:1974se,Weinberg:1978kz,Gasser:1983yg,GSS89}.
The very first  $\chi$PT calculation of nucleon polarizabilities, 
published in 1991 by Bernard, Kaiser and Mei{\ss}ner~\cite{Bernard:1991rq}, quotes
the result shown in the $\cO(p^3)$ column of Table~\tabref{albe}. In the same column, the
numbers in brackets show the result of the so-called heavy-baryon (HB) expansion~\cite{JeM91a}.
Here it means that one additionally expands the full result~\cite{Bernard:1991rq} in powers
of $m_\pi/M_N$, the ratio of the pion and nucleon masses, and drops all but the leading terms
(cf.\ Appendix A).
The $\cO(p^3)$ HB$\chi$PT result thus corresponds to the static nucleon approximation.
The relativistic effects are systematically included in HB$\chi$PT at higher orders, 
but nonetheless even leading order HB$\chi$PT result is widely considered
to be more consistent than the B$\chi$PT (i.e., fully relativistic) one.
The reason for that is that the full relativistic evaluation of the chiral loops may yield
contributions which are of lower order than is given by the power-counting argument.
This ``pathology'', however, does not arise in the case of polarizabilities at $\cO(p^3)$, so as far
as power counting is concerned, the B$\chi$PT result is as good here as the one of HB$\chi$PT.
But even more generally, chiral symmetry ensures that the power-counting violating terms 
to be always accompanied by low-energy constants (LECs), 
hence they can simply be removed in the course of renormalization
of those LECs~\cite{Gegelia:1999gf,Fuchs:2003qc}. 
In simpler terms, there is no problem with power counting in B$\chi$PT.

The present state-of-the-art $\chi$PT studies based on pion and nucleon
degrees of freedom~\cite{McGovern:2001dd,Beane:2004ra} utilize the HB expansion.
They find, however, that despite the 
very reasonable values for polarizabilities, the $\cO(p^3)$ and even $\cO(p^4)$ results
for the Compton-scattering cross sections show significant discrepancy with experimental data
starting from energies of about 120~MeV, especially at backward kinematics. 
The inclusion of the $\Delta$(1232)-resonance as an explicit degree of freedom helps to remedy this discrepancy in the cross sections~\cite{Pascalutsa:2003zk,Hildebrandt:2003fm}. However, it comes at an expense of
a large  contribution to the polarizabilities \cite{Hemmert:1996rw}. This $\Delta$-contribution
is highly unwanted in HB$\chi$PT, since polarizabilties
come out nearly perfect already in the theory without the $\Delta$ (cf.\ the numbers in brackets
in Table~\tabref{albe}).
There is no natural solution to this problem.
One is bound to either omit some of the $\Delta$ contributions
by ``demoting" them to higher orders~\cite{Pascalutsa:2003zk}, or cancel them by ``promoting"
some of the low-energy constants (LECs) to lower orders~\cite{Hildebrandt:2003fm}.

Such an apparent failure of $\chi$PT is sometimes attributed to 
certain ``$\sigma$-meson'' contributions~\cite{Schumacher:2007xr}, 
which $\chi$PT misses. Of course, while the $\si$-meson of the linear sigma model is included
in $\chi$PT, the contribution from the $f_0$(600) is not, but it is doubtful that the $f_0$
can explain it; its two-photon coupling is too small.
 Alternatively, studies based on 
 dispersion relations suggest that some essentially relativistic effects, 
discarded in HB$\chi$PT  as being higher order, are in fact important
because of the proximity of cuts in both pion mass and energy~\cite{Lvov:1993ex,Holstein:2005db,Pascalutsa:2004wm}. 

In our present study we verify the latter scenario and 
perform the calculations in a manifestly Lorentz-covariant fashion,
refraining from the use of the heavy-baryon formalism.
The HB$\chi$PT results can then be recovered by simply expanding
in powers of  pion mass over the baryon mass, $m_\pi/M_B$. 
We thus are coming back to the original (relativistic quantum field theory) ways~\cite{Bernard:1991rq}.
The difference with the original
work~\cite{Bernard:1991rq} is that we compute the Compton scattering 
observables, not only the scalar polarizabilities, and that we include the $\De(1232)$ in addition
to the pion and nucleon degrees of freedom. 

\begin{center}
\begin{table}[t]
\begin{tabular}{||c|c|c|c||c||}
\hline\hline
&  
\multicolumn{3}{|c||}{B$\chi$PT  (HB$\chi$PT)}& PDG \\
\cline{2-4} 
& $ \cO(p^3)$   &$ \cO( p^3)+ \cO(p^4/\vDe) $& $\cO(p^4) $ est. & \cite{PDG2006} \\
\hline
$\alpha^{(p)}$   & 6.8 (12.2) & 10.8 (20.8) & $\pm 0.7$ &  $12.0 \pm 0.6$\\
$\beta^{(p)}$ & $-1.8$ (1.2) & 4.0 (14.7) & $\pm 0.7$ & $1.9\pm 0.5 $ \\ 
\hline\hline
\end{tabular}
\caption{Predictions of baryon $\chi$PT for 
electric ($\al$) and magnetic ($\be$) polarizabilities
of the proton in units of $10^{-4}\,$fm$^3$, compared with 
the Particle Data Group summary of experimental values.
}
\tablab{albe}
\end{table}
\end{center}

Table~\tabref{albe} shows the results of both manifest-covariant 
and HB calculations at all the ``predictive'' orders, i.e., 
below  $\cO(p^4)$ --- the order at which the unknown LECs start to enter.
 A natural estimate of the $\cO(p^4)$
contribution, given in the corresponding column, can serve as an error bar on
the $\chi$PT prediction. A detailed discussion of these results can be
found in Sect.~\secref{polza}. It can be noted, however, how significant
the differences are between the exact and the HB results. 
This is of course not the first and only example where 
B$\chi$PT and HB$\chi$PT are in dissent, see e.g., the case of $\ga N\to \Delta$ transition
\cite{Gail:2005gz,Pascalutsa:2005ts}, or the baryon magnetic moments in 
SU(3)~\cite{Geng:2008mf,Camalich:2009uf}. These differences can often be
significantly diminished by slight improvements of the HB calculations, such as readjusting
the position of the thresholds to have them in the exactly correct place \cite{McGovern:2001dd}.
It is not yet clear, however, how to systematically derive such improvements from the HB formalism itself.

As to why the orders considered here are {\it predictive}, any chiral power-counting scheme will 
tell us that the expansion of the Compton amplitude 
begins at order $p^2$, and that $p^4$ is the order where the
first unknown LECs should enter. In between there are $p^3$
and the $\De$-excitation effects. The counting for the latter
is itself a subject of controversy related to the issue of how to
count the $\De$-nucleon mass difference: $\vDe= M_\Delta-M_N\approx 300$ MeV.
In the  hierarchy  of chiral symmetry breaking  scales, 
$\vDe$ is neither as light as the scale of explicit  symmetry-breaking,
$m_\pi\sim 150$ MeV, nor 
as heavy as the scale of spontaneous symmetry-breaking, $4\pi f_\pi\sim 1$ GeV. 
We treat $\vDe$
 as an independent light scale with the  power-counting
rules defined in Sect.~\secref{lagandpc}. In any case, the leading $\Delta$ effects come before $p^4$.

To recapitulate, in this work we compute the contributions to Compton amplitude up to,  but not including, $\cO(p^4)$
in B$\chi$PT with $\Delta$'s. This is a complete next-to-next-to-leading order (NNLO)
calculation which is entirely expressed in terms of only known LECs.
The details of these calculations are given in Sect.~\secref{techn}.
Polarizabilities and their chiral behaviors are discussed in 
Sect.~\secref{polza} while the results for observables are shown in Sect.~\secref{results}.

Some of these results have recently been reported  in a letter~\cite{Lensky:2008re}.
The present paper is more comprehensive and self-contained.

\section{Chiral Lagrangians and power counting}
\seclab{lagandpc}

The method of constructing the chiral SU(2)  Lagrangians with pion and nucleon fields 
is well known~\cite{Gasser:1983yg,GSS89,Weinberg:1995mt}, and the inclusion of the 
$\Delta$-isobar fields in a Lorentz-covariant fashion has recently been
reviewed~\cite{Pascalutsa:2006up}. We shall list here only the terms relevant
to the present work. The strong-interaction piece is given by
\begin{subequations}
\bea
\lag^{(2)}_\pi &=& \frac{f^2}{4}\, \mathrm{tr} \big( \pa^\mu U \pa_\mu U^\dagger + 2B_0(U M^\dagger + M U^\dagger)
\big), \\
\lag^{(1)}_N &=& \ol N\,\big( i \slad -{M}_{N}  
+ {/\!\!\!v}+  g_A \,  
\slaa\,\ga_5 \big)\, N,
\eqlab{Nlagran}
\\
\eqlab{freeRS}
\lag^{(1)}_{\Delta} &=& \ol\De_\mu \left(i\ga^{\mu\nu\la}\,\pa_\la - 
M_\De\,\ga^{\mu\nu}\right) \De_\nu +  \frac{h_A}{2M_\De}
\left[
i \ol N\, T_a
 \,\ga^{\mu\nu\la}\, (\pa_\mu \De_\nu)\, \mathrm{tr}(a_\la \tau^a )
+ \mbox{H.c.}\right],
\eea  
\end{subequations}
where $U$ is the $SU(2)$ pion field in the exponential parameterization: $U=\exp(i\pi^a\tau^a/f)$,
$f$ is the pion decay constant in the chiral limit,
$M$ is the mass matrix of light quarks, and $B_0$ is a proportionality factor that can be related with
the value of light quark condensate~\cite{Gasser:1983yg}.
In turn, $N$ denotes the isodoublet Dirac field of the nucleon,
$M_N$ is the nucleon mass, and $g_A$ is the axial-coupling 
constant, both taken at their chiral-limit value, and
 the vector and axial-vector chiral fields above are defined 
in terms of the pion field, $\pi^a(x)$, as
\begin{subequations}
\bea
v_\mu & \equiv & \half\, \tau^a v_\mu^a(x) = \frac{1}{2i} \left(u \,
\pa_\mu u^\dagger+u^\dagger \pa_\mu  u \right), \\
a_\mu & \equiv & \half \,\tau^a a^{\,a}_\mu(x) =
  \frac{1}{2i} \left(u^\dagger \,
\pa_\mu  u- u \,\pa_\mu  u^\dagger \right), 
\eea%
\eqlab{currents}%
\end{subequations}%
where $u=\exp(i\pi^a \tau^a/2f )=U^{1/2}$. Finally, $\Delta_\nu$ is the Delta isobar 
Rarita--Schwinger field with mass $M_\Delta$, and $h_A$ is the $\pi N\Delta$ coupling constant
whose value is fixed to the $\Delta\to\pi N$ decay width of $115$ MeV. The
antisymmetrized products of Dirac matrices in the above equations are defined as: $\gamma^{\mu\nu}=\frac{1}{2}(\gamma^\mu\gamma^\nu-\gamma^\nu\gamma^\mu)$ and
$\gamma^{\mu\nu\lambda}=\frac{1}{2}(\gamma^{\mu\nu}\gamma^\lambda+\gamma^\lambda\gamma^{\mu\nu})$. The isospin $1/2\to 3/2$ transition
matrix $T$ is normalized such that $T^aT^{b\,\dagger}=\frac{1}{3}(2\delta^{ab}-i\epsilon^{abc}\tau^c)$.

The electromagnetic interaction is added as usual through the minimal 
substitution:
\begin{subequations}
\bea
\pa_\mu N &\to & \pa_\mu N -  i e A_\mu\half(1+\tau_3) N ,\\
\pa_\mu \pi^a &\to & \pa_\mu \pi^a - eA_\mu\eps^{ab3}\pi^b\,,
\eea
\end{subequations}
where $A_\mu$ is the photon field. The minimal coupling of the photon to the Delta field gives contributions to Compton scattering which are of higher orders than the ones considered in this work.

There is as well a number of nonminimal terms:
\begin{subequations}
\bea
\lag^{(2)}_N &=&  \frac{e\kappa_{p,n}}{4M_N} 
\,\ol N\,\half (1\pm \tau_3) \,\ga^{\mu\nu}\, N \,F_{\mu\nu} , \\
\lag^{(2)}_\Delta &=&  \frac{3e}{2M_N(M_N+M_\Delta)}\,\ol N\, T_3
\left(i g_M  \tilde F^{\mu\nu}
- g_E \gamma_5 F^{\mu\nu}\right)\,\pa_{\mu}\De_\nu
+ \mbox{H.c.},
\\
\lag^{(4)}_{\mathrm{WZW}}&=&  -\frac{e^2 }{32\pi^2 f} F_{\mu\nu} \tilde F^{\mu\nu}\pi_3 \, .\nn
\eea
\end{subequations}
Here, $F^{\mu\nu}$ and $\tilde F^{\mu\nu}$ are the photon field strength tensor and its dual tensor defined
as $F^{\mu\nu}=\pa^\mu A^\nu-\pa^\nu A^\mu$, $\tilde F^{\mu\nu}=\frac{1}{2}\epsilon^{\mu\nu\rho\lambda}F_{\rho\lambda}$;
$\kappa_p$ ($\kappa_n$) stands for the proton's (neutron's) anomalous magnetic moment; 
$g_E$ and $g_M$ are $\gamma N\Delta$ electric and magnetic couplings, respectively,
which are well known from the 
analysis of pion-photoproduction $P_{33}$ multipoles~\cite{Pascalutsa:2005ts}.
Here we differ from the strategy adopted in Ref.~\cite{Pascalutsa:2003zk}, 
where, in the absence of any $\chi$PT
analysis of pion-photoproduction at the time, the values of $g_E$ and $g_M$ were fitted
to Compton scattering data, with a rather unsatisfactory result. 
The precise values of all the parameters used in the present work are given in Table~\tabref{params}.

\begin{table}[tb]
\begin{tabular}{||c| l ||}
\hline\hline
 & \\
$ \cO(p^2)$ &\,  $\frac{e^2}{4\pi} = \frac{1}{137} $, $M_N = 938.3$ MeV, 
$\hbar c = 197$ MeV$\cdot\,$fm\\
&\\
$ \cO(p^3)$ & \, $g_A=1.267$, $f_\pi = 92.4$ MeV, $m_\pi = 139$ MeV, $m_{\pi^0} = 136$ MeV, $\kappa_p=1.79$\, \\ 
& \\
 \, $ \cO(p^4/\vDe)  \, $& \, $M_\De= 1232$ MeV, $h_A=2.85$, $g_M = 2.97$, $g_E=-1.0$\\
& \\
$\cO(p^4) $  & \, $\al_0, \be_0 = \pm  \frac{e^2}{4\pi M_N^3}$\, \\
& \\
\hline\hline
\end{tabular}
\caption{Parameters  (fundamental and low-energy constants) at the order they first appear.}
\tablab{params}
\end{table}

Inclusion of the $\Delta$-isobar fields in a Lorentz-covariant fashion raises the
consistency problems of higher-spin field theory~\cite{Johnson:1961vt,Velo:1969bt,Piccinini:1984dd}. 
 The $\Delta$-isobar couplings used here possess the property of invariance
under a gauge transformation: $\Delta_\mu\to \Delta_\mu + \partial_\mu \epsilon$, where
$\epsilon$ is an arbitrary spinor field. This ensures the decoupling of unphysical spin-$1/2$
degrees of freedom and eliminates 
the consistency problems~\cite{Pascalutsa:1998pw,Pascalutsa:1999zz}. 
It is far  less  straightforward to reconcile this extra gauge symmetry with
other symmetries of the chiral Lagrangian. For recent progress see 
Refs.~\cite{Pascalutsa:2006up,Pascalutsa:2000kd,Krebs:2008zb}.

Coming to the power counting, for the pion and nucleon contributions we shall use the
usual scheme~\cite{GSS89}, i.e., a graph with $V_k$ vertices from $\lag^{(k)}$, 
$L$ loops, $N_\pi$ pion and $N_N$ nucleon lines is of order $p^n$ with  
\beq
n = \sum_k  k V_k + 4L - 2N_\pi -N_N\, .
\eeq
In the case of Compton scattering by $p$ we understand the photon energy or/and the pion mass
as compared with $4\pi f_\pi \sim 1$ GeV.

The graphs with $\Delta$s are more tricky because for small $p$ they go as 
\beq
S_\De \sim \frac{1}{p\pm \vDe}
\eeq
rather than simply $1/p$ as the nucleon propagators. The new scale 
\beq
\vDe = M_\De - M_N \simeq 293 \, \mbox{MeV}
\eeq
is neither as light as $m_\pi$ nor as as heavy as $4\pi f_\pi$, hence can and will
 be treated independently.
For energies comparable to the pion mass we choose to additionally expand in $p/\vDe$, and hence
the $\Delta$ propagator counts as $1/\vDe$, while a graph with $N_\De$ internal lines
contributes to order 
\beq
p^n \left(\frac{1}{\vDe}\right)^{N_\De}\,.
\eeq
 For definiteness, when needed, we count $\vDe^2$ to be of  $\cO(p)$, i.e., 
the ``$\de$ counting" scheme \cite{Pascalutsa:2003zk}.

For the power counting in the region where the energies
are of order of $\vDe$ (the resonance region) see~\cite{Pascalutsa:2003zk,Pascalutsa:2005ts}.
Hereby we limit ourselves to the low-energy region, 
\beq
p\sim m_\pi \ll\vDe\ll 4\pi f_\pi .
\eeq

\section{Compton amplitude at NNLO}
\seclab{techn}

\subsection{Graphs and the nucleon field redefinition}

The chiral expansion for the Compton amplitude begins with graph $(1)$ in \Figref{born}
and its crossed counterpart. Nominally they both are of $\cO(p)$, however, together they
simply give the Thomson amplitude, which is of $\cO(p^2)$. We thus refer
to $\cO(p^2)$ as the leading order (LO). The other graphs in \Figref{born}
contribute to $\cO(p^3)$, the next-to-leading order (NLO). 

\begin{figure}[tb]
\centerline{\epsfclipon   \epsfxsize=14cm%
  \epsffile{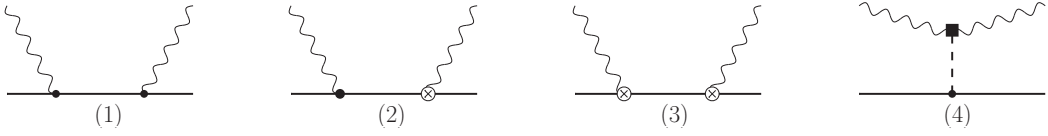} 
}
\caption{
The tree graphs evaluated in this work. Graphs obtained from these by
crossing and time reversal are not shown, but are evaluated too. Dots stand for
the (leading) $\gamma NN$ and $\pi NN$ vertices from $\lag^{(1)}$, whereas crossed circle denotes
the $\lag^{(2)}$ coupling via anomalous magnetic moment. Filled square stands
for the WZW-anomaly $\pi^0\gamma\gamma$ vertex from $\lag^{(4)}$.
}
\figlab{born}
\end{figure}

\begin{figure}[bt]
\centerline{\epsfclipon   \epsfxsize=13cm%
  \epsffile{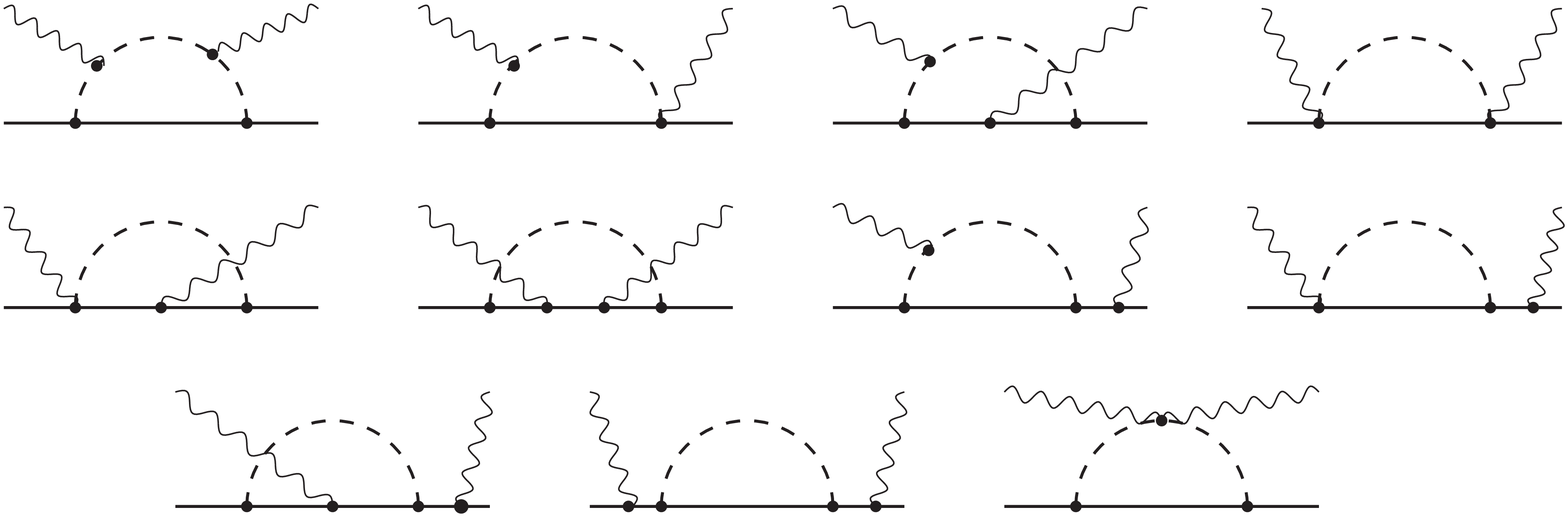} 
}
\caption{
The one-loop graphs contributing to Compton scattering at $\cO(p^3)$. Graphs obtained from these by
crossing and time reversal are not shown. 
}
\figlab{loops_pv}
\end{figure}

At NLO we also have the one-loop contributions shown in \Figref{loops_pv},
but before evaluating them we make a redefinition of the nucleon field, $N\to \xi N$, 
where
\beq
\xi = \exp\left(\frac{ig_A\,\pi^a \tau^a}{2f}\ga_5 \right)\,.
\eeq
The first-order chiral Lagrangian \Eqref{Nlagran} then becomes:\footnote{In our conventions
$\ga_5^\dagger= \gamma_5$, hence 
$\xi^\dagger=\exp(-ig\pi^a \tau^a\ga_5/2f_\pi)$, $\xi \xi^\dagger =1$.
Note also that $\ol N\to \ol N \xi$, and $\xi\ga^\mu\xi =\ga^\mu$.}
\bea
\eqlab{Nlagran2}
{\lag_N'}^{(1)}& = & \ol N\,\xi\,( i \sla{D} -{M}_{N} +  g_A \,  
\slaa\,\ga_5 )\, \xi\,N\,\nn\\
&=& \ol N\,( i \slad -{M}_{N} )\, N + M_N \,\ol N \,(1-\xi^2) N \\
 && + \ol N \, (  \xi \,i \slad\,\xi - \xi\,{v\!\!\! \slash} \,\xi 
+  g_A \, \xi\, \slaa\,\ga_5\,\xi ) \, N \,.\nn
\eea

The two Lagrangians 
are equivalent, in the sense of equivalence theorem, however, 
may have drastically different forms when
expanded in the pion field. For the one-loop contributions to Compton scattering
it is sufficient to expand up to the second
order in the pion field: 
\begin{subequations}
\bea
v_\mu &=&  \frac{1}{4f^2}\, \tau^a\veps^{abc} 
\pi^b\,\pa_\mu\pi^c+ \cO(\pi^3),\\
a_\mu &=& \frac{1}{2f}\tau^a\pa_\mu\pi^a+ \cO(\pi^3),\,\\
\xi &=& 1+ \frac{i g_A}{2f} \tau^a \pi^a \ga_5 - \frac{g_A^2}{8f^2}\pi^2 
+ \cO(\pi^3).
\eea
\end{subequations}
The original and the redefined Lagrangians take, respectively,
the following form:
\bea
\lag^{(1)}_N & = & \ol N\,\left( i \slad -{M}_{N}
+ \frac{g_A}{2f} \tau^a \slad\,\pi^a\ga_5 \right. \nn\\
&& - \left. \frac{1}{4f^2} \, \tau^a \veps^{abc}  \pi^b\,\slad \,\pi^c
\right)\, N  + \cO(\pi^3) \,,
\eqlab{expNlagran}\\
{\lag'}_N^{(1)} & = &  \ol N\,\left( i \slad -{M}_{N}\, 
-\, 
i\, \frac{ g_A}{f} M_N \tau^a\pi^a\ga_5 +  \frac{g_A^2}{2f^2} M_N \pi^2
\right. \nn\\
&& \left. -\frac{(g_A-1)^2}{4f^2} \, \tau^a\veps^{abc}  \pi^b\,\slad \,\pi^c
\right)\, N\,
 + \cO(\pi^3)\,.
\eqlab{expNlagran2}
\eea
The major difference between the two forms 
is that the pseudovector $\pi NN$ coupling is transformed into a pseudoscalar one, while
the Weinberg--Tomozawa $\pi\pi NN$ term, which resembles a $\rho$-meson exchange,
gets replaced by an isoscalar term akin to the remains of an integrated-out $\sigma$-meson
in the linear $\sigma$ model. The isovector $\pi\pi NN$ term, which is now proportional to $(g_A-1)^2$, does not
give any contribution to Compton amplitude at one-loop level.

\begin{figure}[tb]
\centerline{\epsfclipon   \epsfxsize=13cm%
  \epsffile{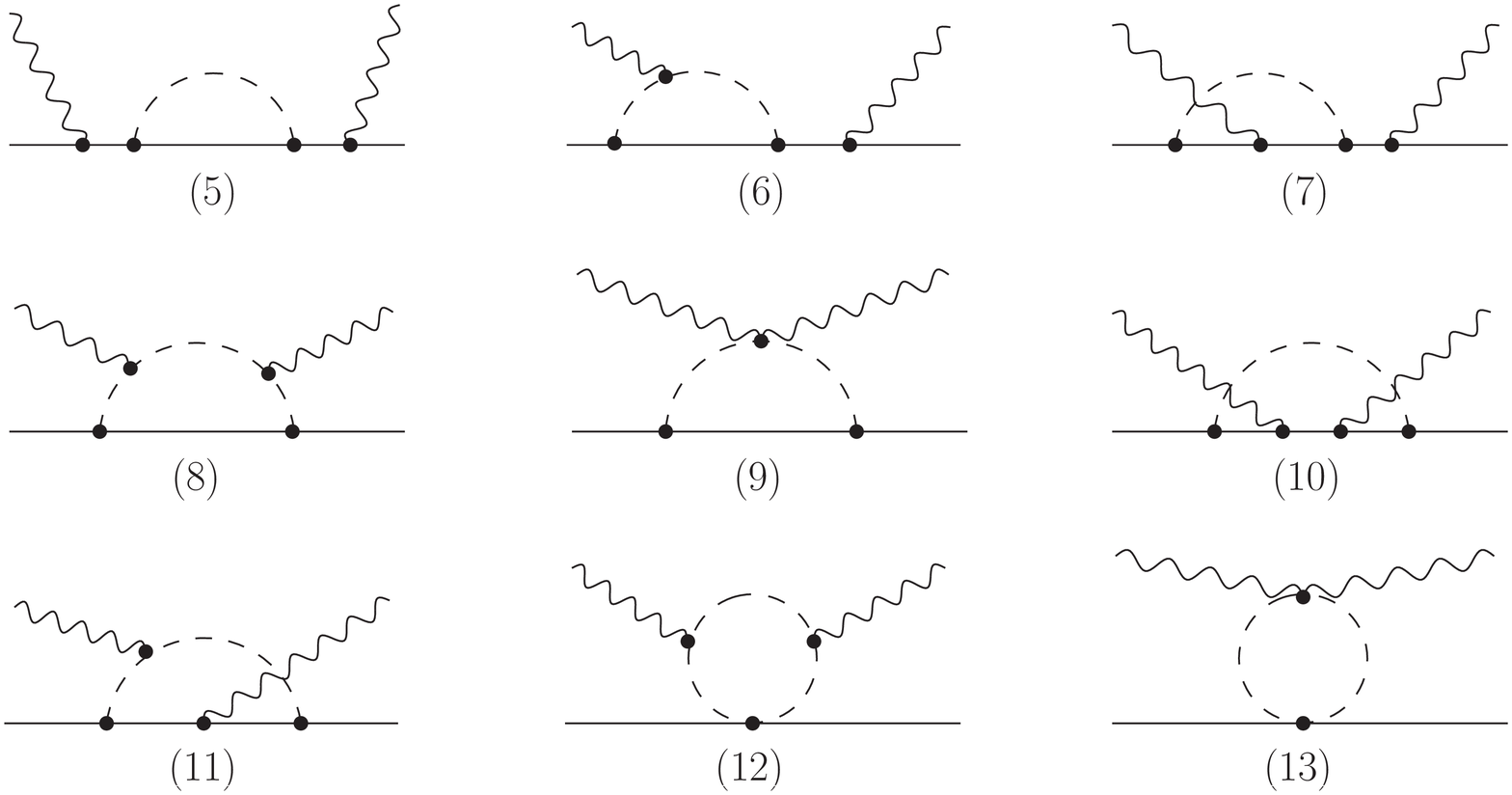} 
}
\caption{
The NLO loop graphs evaluated in this work. Graphs obtained from these by
crossing and time reversal are not shown, but are evaluated too.}
\figlab{loops}
\end{figure}

Also, in the NLO  loops, the photon couples only minimally, i.e., to the electric charge of the
pion and nucleon.
Now that the pion couples to the nucleon via pseudoscalar coupling, there is no Kroll--Ruderman ($\gamma \pi NN$)
term arising, and hence the number of one-loop graphs is
reduced. The resulting expressions for amplitudes become simpler.
 
As a result, the loop graphs shown in \Figref{loops_pv} with couplings from the Lagrangian \Eqref{expNlagran}
transform to the graphs shown in \Figref{loops} with the couplings from \Eqref{expNlagran2}.
We have also checked explicitly that the two sets
of one-loop diagrams give identical expressions for the Compton amplitude.

Although the main purpose of the above field redefinition is to simplify the calculation,
it does give more insight about the chiral dynamics. First of all, 
it explains how Metz and Drechsel~\cite{Metz:1996fn}, calculating
polarizabilities in the linear $\sigma$ model with a heavy $\sigma$-meson,
obtain to one loop exactly the same result as B$\chi$PT at $\cO(p^3)$~\cite{Bernard:1991rq}. 
Secondly, observing that graphs (12) and (13) vanish in the forward kinematics, 
we can see that  chiral symmetry plays less of a role in the forward Compton scattering
at order $p^3$.

\begin{figure}[b]
\centerline{\epsfclipon   \epsfxsize=13cm%
  \epsffile{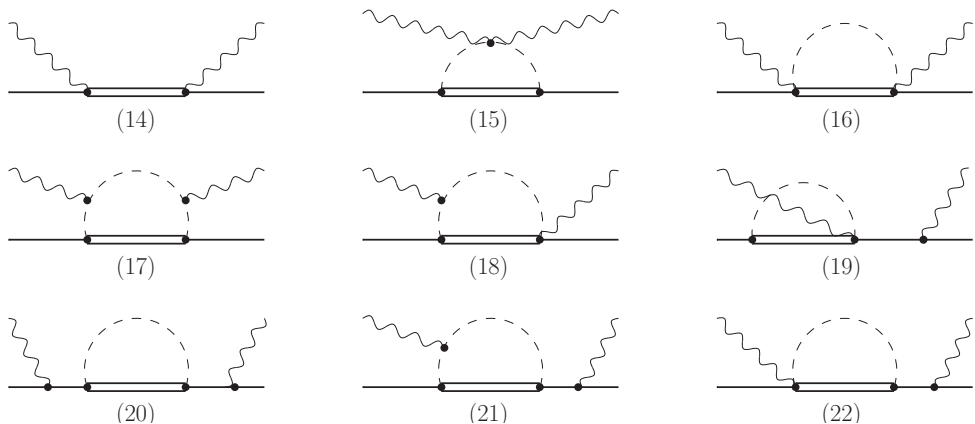} 
}
\caption{
The NNLO graphs evaluated in this work. Double lines denote the propagator of the $\Delta$.
Graphs obtained from these by
crossing and time reversal are not shown, but are evaluated too.}
\figlab{loopsD}
\end{figure}

Going  to $\cO(p^4/\vDe)$ we encounter the graphs with one $\De$-isobar propagator
shown in \Figref{loopsD}.
The nucleon-field redefinition does not affect these contributions at this order.

Note that the graphs where photons couple minimally to $\Delta$ contain more than
one $\Delta$ propagator and therefore should be suppressed by extra powers of
$p/\vDe$. However, their lower-order contributions are important for electromagnetic
gauge invariance and therefore for the renormalization program. In particular,
the lower-order contributions of chiral loops should not affect the result of the low-energy
theorem (LET)~\cite{Low:1954kd},
 and this condition is automatically satisfied for a subclass of graphs which
obeys gauge invariance. The loop graphs in \Figref{loopsD}
form such a subclass for the case of neutral $\Delta$. In reality the $\De$ comes in
four charge states (isospin $3/2$), and hence a gauge invariant set will
in addition have the higher-order graphs where
photon couples minimally to the $\De$.
To make the subclass of loop graphs 
in \Figref{loopsD} gauge invariant without the higher-order graphs,
we used the following procedure:
\begin{itemize}
\item[---] The one-particle-irreducible (1PI) graphs, \Figref{loopsD}(15--18) are computed with
the correct isospin factors, i.e., summing over all charge states of the $\De$. The
isospin factors for the one-particle reducible (1PR) graphs (19--22) are chosen
such that their ratio to the isospin factors of 1PI graphs is the same as in the neutral $\De$ case.
\end{itemize}
This procedure automatically ensures exact gauge invariance and thus effectively includes
the lower-order contributions of the one-loop graphs with minimal coupling of photons to the $\De$. In case when the latter graphs are included explicitly, the isospin factors of 1PR
graphs can be restored to actual values. This, however, will not affect
the result at the order considered here.

The graphs in \Figref{loops} and in \Figref{loopsD} were eventually computed by us with the help of
the  symbolic manipulation tool FORM \cite{FORM} and the 
LoopTools library~\cite{Hahn:2006qw} using dimensional regularization.

\subsection{Renormalization}

In accordance with the LET~\cite{Low:1954kd},
the loop contributions shown in \Figref{loops} and in \Figref{loopsD} may contribute
to the renormalization of nucleon mass, field, charge, and anomalous magnetic moment.
We have adopted the on-mass-shell renormalization scheme, 
not the extended on-mass-shell renormalization (EOMS)~\cite{Fuchs:2003qc}. 
The difference is that in EOMS the above-listed quantities are taken to be at their chiral limit value
while here we simply take the values at the actual pion mass, see, e.g., Table~\tabref{params}. 

Let us discuss first the contributions to nucleon self-energy and $\gamma NN$ vertex
corresponding to the nucleon loops (5--7) in \Figref{loops}. The corresponding amputated
diagrams give contributions to nucleon self-energy (5) and to $\gamma NN$ vertex (6), (7).
More specifically, the contributions to the self-energy and the $\gamma NN$ vertex can be written in the
following form:
\begin{eqnarray}
i\Sigma(\cancel{p})&=&i\Sigma(M_N)+i(\cancel{p}-M_N)\Sigma^\prime(M_N)\nonumber\\
 & &+S(\cancel{p}-M_N),\\
i\Gamma^\mu (p,p_s)&=&i\gamma^\mu F_1(p_s^2)-i\gamma^{\mu\nu}q_\nu F_2(p_s^2)+i\cancel{q}p^\mu F_3(p_s^2)\nonumber\\
& &+i(\cancel{p}_s-M_N)\gamma^\mu F_4(p_s^2),
\end{eqnarray}
where $p$ is the initial nucleon momentum, $q$ is the initial photon momentum, and $p_s=p+q$.
The function $S$ is finite and its expansion in powers of $\cancel{p}-M_N$ starts from a quadratic term. Of all the
functions $F_1\dots F_4$ that contribute to the $\gamma NN$ vertex only $F_1$ is divergent. It contributes
to the renormalization of charge. 
Function $F_2$ contributes to a renormalization of the nucleon's
anomalous magnetic moment. In this case the renormalization is finite \cite{Holstein:2005db}.

After the renormalization, the self-energy and the $\gamma NN$ vertex
can be written in the following form:
\begin{eqnarray}
i\Sigma_R(\cancel{p})&=&i\Sigma(\cancel{p})-i\Sigma(M_N)-i(\cancel{p}-M_N)\Sigma^\prime(M_N)\nonumber\\
&=&S(\cancel{p}-M_N)\\
i\Gamma_R^\mu(p,p_s)&=&i\gamma^\mu \overline{F}_1(p_s^2)-i\gamma^{\mu\nu}q_\nu \overline{F}_2(p_s^2)
+i\cancel{q}p^\mu F_3(p_s^2)\nonumber\\
& &+i(\cancel{p}_s-M_N)\gamma^\mu F_4(p_s^2),
\end{eqnarray}
where $\overline{F}_{1,2}(p_s^2)=F_{1,2}(p_s^2)-F_{1,2}(M_N^2)$ are subtracted functions. Note that
functions $F_3(p_s^2)$ and $F_4(p_s^2)$ do not get subtracted; indeed, the Lorentz structures
that correspond to these functions are purely off-shell --- they give zero when both nucleons are on-shell
(i.e., when both $p$ and $p_s$ are on-shell momenta),
so they do not contribute to the renormalization of charge or magnetic moment. However,
both these functions play an important role in making the complete Compton scattering amplitude
gauge invariant. Note also the fact that after the renormalization of nucleon mass, wave function, charge, and anomalous magnetic moment is performed, the remaining expressions for the nucleon loops (5--7) become finite.

Now we come to the loops with $\Delta$,  \Figref{loopsD}. The corresponding amputated loops also give contributions to nucleon self-energy and to $\gamma NN$ vertex, and the corresponding expressions can be written
in a full analogy to the case of nucleon loops. The loop (19) in \Figref{loopsD} also gives a contribution
to $\gamma NN$, however, this contribution is fully off-shell and momentum-independent:
\begin{eqnarray}
i\Gamma^\mu (p,p_s)&=&i\cancel{q}p^\mu A+i(\cancel{p}_s-M_N)\gamma^\mu B,
\end{eqnarray}
where $A$ and $B$ are constants. Nevertheless, it is important to take this contribution into account in order to preserve the electromagnetic gauge invariance.
The renormalization of nucleon self-energy and $\gamma NN$ vertex proceeds for these loops
in complete analogy to the purely nucleon loops. 

In the case of $\Delta$ loops, however,  we obtain in addition some higher-order 
divergences, i.e., ultraviolet divergences of ${\cO}(p^4)$.
They are to be renormalized by a corresponding ${\cO}(p^4)$ contact term. At this stage
it is customary to use the $\overline{MS}$ subtraction of the higher-order divergences,
see e.g., Ref.~\cite{Kubis:2000zd}. 
We have implemented the $\overline{MS}$ scheme for the higher-order divergences
by putting the dimreg factor  equal to zero (see Appendix A for more
detail).

\subsection{Consistency with forward-scattering sum rules}

The dispersion relations enjoy a special role in nucleon Compton scattering,
see Ref.~\cite{Drechsel:2002ar} for a review. First of all, 
practically all up-to-date empirical values of nucleon polarizabilities 
are extracted from data with
the use of a model based on dispersion relations~\cite{Lvov:1980wp,L'vov:1996xd}.
Secondly, in the forward kinematics, the Compton amplitude can be related to 
an integral over energy of the photoabsorption cross section, which in combination
with the low-energy expansion yields a number of model-independent sum rules.
A famous example is the Baldin sum rule:
\beq
\alpha+\beta=\frac{1}{2\pi^2}\int\limits^\infty_0
 d\nu \frac{\sigma_{\mathrm{tot}}(\nu)}{\nu^{2}-i0}\eqlab{BSR},
\eeq
where the sum of polarizabilities is related
to an integral of the total photoabsorption cross section $\sigma_{\mathrm{tot}}$ over the photon lab-frame energy $\nu$.

In general, the forward Compton-scattering amplitude can be 
decomposed into two scalar functions of a single variable in the following way:
\beq
T_{fi}(\nu)=\vec\epsilon^{\,\prime*}\cdot\vec\epsilon \,f(\nu)
+i\vec\sigma\cdot(\vec\epsilon^{\,\prime*}\times\vec\epsilon\,)\,\nu\,g(\nu),
\eeq
where $\vec\epsilon^{\,\prime},\ \vec\epsilon$ are the polarization vectors of the initial and final photons, respectively,
and $\vec\sigma$ are the Pauli spin matrices. The functions $f$ and $g$ are even functions of $\nu$. 
Using analyticity and  the optical theorem, one can write down the following 
sum rules:
\begin{subequations}
\eqlab{sumrules}
\bea
f(\nu) &=& f(0)+\frac{\nu^2}{2\pi^2}\int\limits^\infty_0
 d\nu^\prime\frac{\sigma_{\mathrm{tot}}(\nu^\prime)}{\nu^{\prime\,2}-\nu^2-i0}\eqlab{fdis}\,,\\
g(\nu) &=& \frac{1}{4\pi^2}\int\limits^\infty_0
 d\nu^\prime\,\nu^\prime\, 
\frac{\sigma_{1/2}(\nu^\prime)-\sigma_{3/2}(\nu^\prime)}{\nu^{\prime\,2}-\nu^2-i0}\eqlab{gdis}\,,
\eea
\end{subequations}
where $f(0) = -e^2/M_N$ is the Thomson amplitude and $\sigma$ is
the doubly polarized photoabsorption cross section, with the index indicating the helicity
of the initial photon--nucleon state; $\sigma_{\mathrm{tot}} = \half ( \sigma_{1/2}+\sigma_{3/2})$.

These sum rules should also hold for the individual contributions of the
loop graphs in \Figref{loops}. In this case 
the photoabsorption process is given by the Born graphs of  single-pion
photoproduction, for which analytic expressions 
exist~\cite{Holstein:2005db,Pascalutsa:2004ga}:
\bea
\si_{1/2}^{(\pi^0 p)}+\si_{3/2}^{(\pi^0 p)} &=&
\frac{\pi C}{M_N\nu^3}\,\left\{ [\nu^2-\mu^2 x_N s]\,
\ln\frac{x_N+\la}{x_N-\la} + 2 \la \left[ \nu^2 (x_N-2)  + s \mu^2  \right]
\right\},  \nn\\
\si_{1/2}^{(\pi^+ n)} + \si_{3/2}^{(\pi^+ n)}&=& \frac{2\pi C}{M_N\nu^3}\,
\left\{ -x_\pi  s \mu^2\ln\frac{x_\pi+\la}{x_\pi-\la} +2\la
\, (x_N\nu^2 +s \mu^2)\right\}, \nn\\
\si^{(\pi^0 p)}_{1/2} - \si^{(\pi^0 p)}_{3/2} &=&
\frac{\pi C}{\nu^2}\,\left\{ - \big(2x_N \frac{s}{M_N^2}+1- \frac{\nu}{M_N}\big) \ln\frac{x_N+\la}{x_N-\la} 
\right. \\
&&\,\,\,\,\left.  +
\, 2 \la\big[ \frac{\nu}{M_N} (x_N-2) + \frac{s}{M_N^2} (x_N+2) \big] \right\},   \nn\\
\si^{(\pi^+ n)}_{1/2}- \si^{(\pi^+ n)}_{3/2} &=& \frac{2\pi C}{\nu^2}\,
\left\{  \mu^2 \ln\frac{x_\pi+\la}{x_\pi -\la} - 2\la\big( \frac{s}{M_N^2} x_\pi -\frac{\nu}{M_N} x_N\big)\right\}
, \nn
\eea
where $C= [e g_A M_N/(4\pi f_\pi) ]^2$, $\mu=m_\pi/M_N$, $s=M_N^2 +2M_N\nu$, and 
 \bea
&& x_N=(s+M_N^2-m_\pi^2)/2s,\nn\\
&& x_\pi=(s-M_N^2+m_\pi^2)/2s ,\\
&& \la = (1/2s)\sqrt{s-(M_N+m_\pi)^2}\sqrt{s-(M_N-m_\pi)^2} ,\nn
\eea
are the fractions of nucleon and pion energy ($x$) and momentum ($\la$) in the center-of-mass frame.


We have verified indeed that 
the (renormalized) $p^3$ loop contributions in \Figref{loops} 
fulfill  the sum rules in \Eqref{sumrules} exactly for any positive $\nu$.

It is interesting to note that the leading-order pion photoproduction amplitude,
which enters on the right-hand side of \Eqref{sumrules}, is independent of
whether one uses pseudovector or pseudoscalar $\pi NN$ coupling~\cite{Pascalutsa:2004ga}. 
It essentially means that chiral symmetry of the effective Lagrangian plays no role 
at this order. 
The latter statement can, by means of the sum rule, be extended 
to the forward Compton amplitude
at $\cO(p^3)$. On the other hand, the graphs $(12)$ and $(13)$ in \Figref{loops},
being the only ones beyond the pseudoscalar theory, take the sole role
of chiral symmetry. In the forward kinematics these graphs indeed vanish
but play an important role in the backward angles. Without them the values of $\al$ and $\be$ would be entirely different.
The value of $\al+\be$ would of course be the same, but $\al -\be$
would (approximately) flip sign. Furthermore, in the chiral limit, the value of
$\al -\be$ would diverge as $1/m_\pi^{2}$ (instead of $1/m_\pi$ as it should). 
We thus arrive at the conclusion that chiral symmetry of the effective Lagrangian
plays a more prominent
role in backward Compton scattering.

\subsection{Error due to $\cO(p^4)$ effects}
In the previous publication~\cite{Lensky:2008re}
we simply adopted the error estimate from Ref.~\cite{Pascalutsa:2003zk}.
However, in this work we compute to one order higher than in Ref.~\cite{Pascalutsa:2003zk} and
hence the error analysis needs to be revised accordingly. 

An error of an effective-field theory calculation is an estimate of higher-order
effects assuming their natural size. The higher-order effects not included in our calculation begin
at $\cO(p^4)$, i.e., the order at which the polarizability LECs, $\de\al$ and $\de\be$, arise.
The naturalness assumption requires  these constants to be of order of unity in the units of the chiral symmetry breaking scale of a GeV. To be more specific we assume the absolute
value of these constants (in the $\ol{MS}$ scheme) is limited by 
\beq
\al_{(err.)}=\be_{(err.)}=(e^2/4\pi)/M_N^3 \approx 0.7\,  \times\,10^{-4}\,\mathrm{fm}^3,\,\,\,
\eeq
This number gives a natural estimate of the error on polarizability
values we have obtained at NNLO.  It is not difficult to find how this error
propagates to observables, once the effect of the LECs on those observables is known.
For example,  for the unpolarized 
differential cross section,  the error is given by [cf.~\Eqref{unpolcs}]:
\beq
\frac{d\si^{(err.)}}{d\Omega} =  8\, (e^2/4\pi)\,\Phi \,  \nu\nu\,'  \left[(1+z)^2 \al_{(err.)}^2 + 
(2z)^2 \be^2_{(err.)} \right]^{1/2}
\eqlab{error}
\eeq 
 where $z=\cos\th_{\mathrm{lab}}$, $ \nu$ ($\nu'$) the laboratory energy of the incident (scattered)
 photon, and 
 \beq
 \Phi = \left\{  \begin{array}{lc}
\frac{1}{4s} &  \mbox{(center-of-mass frame)}\,, \\
\frac{1}{4M_N^2}\left( \frac{\nu'}{\nu} \right)^2  & \mbox{(lab frame)}\,.
\end{array} \right.
\eqlab{phasespace}
\eeq

Let us emphasize that we do not include the errors due to the uncertainty in the values of  
parameters in Table~\tabref{params} or due to the $\cO(p^5/\vDe^2)$ effects which
stem from graphs with two $\De$ propagators. Our errors are thus
underestimated, however, they can directly serve as an  indicator of  sensitivity to
the polarizability LECs at $\cO(p^4)$.

\section{Proton polarizabilities}
\seclab{polza}

The chiral-loop contribution to scalar polarizabilities of the proton which arise from the NNLO
calculation of the Compton amplitude is given in the Appendix A. In addition, we have 
the tree-level $\De$(1232) contribution from graphs (14) in \Figref{loopsD} and its crossed,
given by~\cite{Pascalutsa:2003zk}:
\begin{eqnarray}
 \alpha \,(\mbox{$\De$-excit.})&=& -\frac{2e^2g_E^2}{4\pi(M_N+M_\Delta)^3}\simeq - 0.1\,,\\
\beta \,(\mbox{$\De$-excit.}) &=& \frac{2e^2g_M^2}{4\pi(M_N+M_\Delta)^2\vDe}\simeq 7.1\,.
\end{eqnarray}
Here and in what follows the numerical values are given in the units of $10^{-4}\,$fm$^3$.

\begin{figure}[tb]
\centerline{\epsfclipon  \epsfxsize=8cm%
  \epsffile{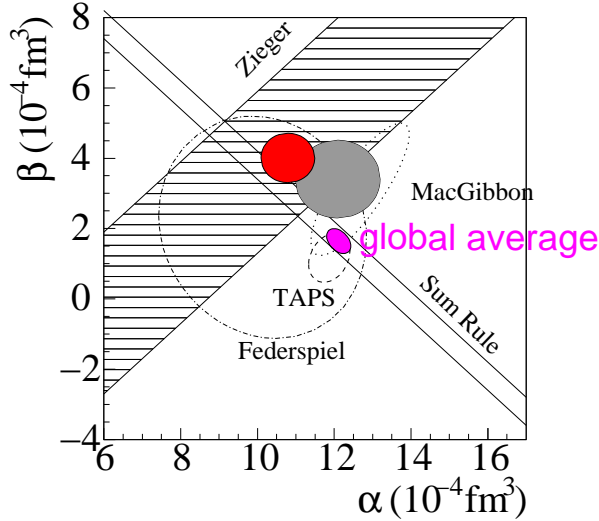} 
}
\caption{(Color online) The scalar polarizabilities of the proton. 
Our result is shown by the red blob. The $\De$-less 
HB$\chi$PT
result~\cite{Beane:2004ra} is shown by the grey blob. Experimental
results are from Federspiel et~al.~\cite{Federspiel:1991yd},
Zieger et al.~\cite{Zieger:1992jq}, MacGibbon et al.~\cite{MacG95},
and TAPS~\cite{MAMI01}.
``Sum Rule'' indicates the Baldin sum rule constraint on $\alpha+\beta$~\cite{Bab98}.
``Global average'' represents the PDG summary~\cite{PDG2006}.} 
\figlab{potato}
\end{figure}

\begin{figure}[tb]
\centerline{\epsfclipon  \epsfxsize=8cm%
  \epsffile{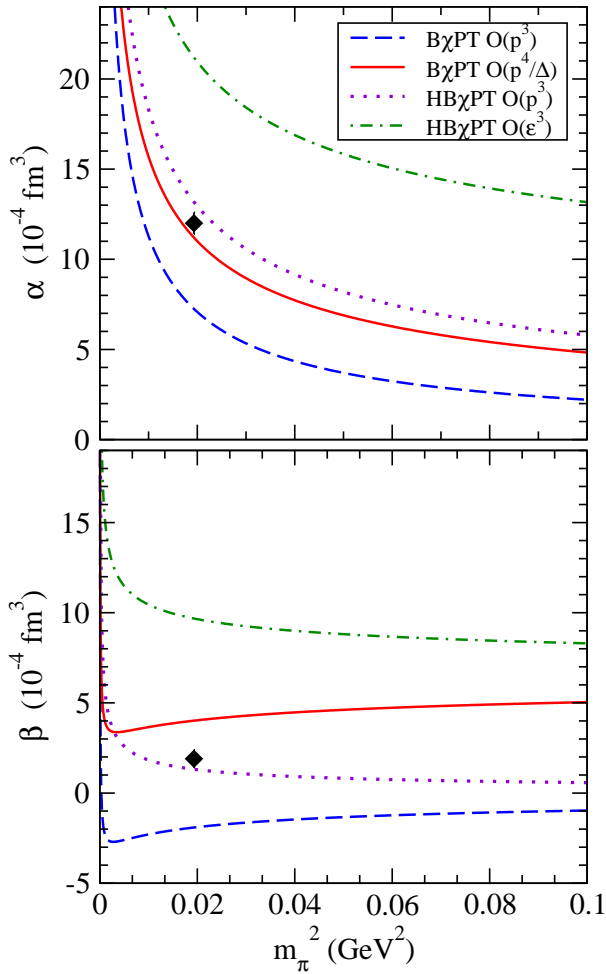} 
}
\caption{(Color online) The pion-mass dependence of proton polarizabilities.
Data points at the physical pion mass represent the PDG values. The legend for the curves
is in the upper panel.} 
\figlab{chiral}
\end{figure}

The numerical composition of the full result thus looks as follows:
\begin{eqnarray}
 \alpha&=& \underbrace{6.8}_{\cO(p^3)} + \underbrace{(-0.1) + 4.1}_{\cO(p^4/\vDe)} = 10.8\,,\\
\beta&=&\underbrace{-1.8}_{\cO(p^3)} +\underbrace{ 7.1-1.3}_{\cO(p^4/\vDe)} =4.0  \,.
\end{eqnarray}
As explained earlier, a natural estimate of $\cO(p^4)$ contributions yields an uncertainty
of at least $\pm 0.7$ on these values. In \Figref{potato} this result, shown by the red
blob, is compared with the empirical information, and with the $\De$-less 
$\cO(p^4)$ HB$\chi$PT
result of Beane et~al.~\cite{Beane:2004ra}.

We can clearly see a few-sigma discrepancy of our result with the TAPS-MAMI
determination of polarizabilities~\cite{MAMI01}. On the other hand, 
as shown in the next section, our result agrees with TAPS data for 
the  Compton differential cross sections. Of course we compare with the data
at the lower energy end (below the pion threshold) where 
polarizabilities play the prominent role.
The extraction of the polarizabilities in Ref.~\cite{MAMI01} has also been influenced by
data above the $\De$-resonance region to which we cannot compare. 
Clearly an extraction of scalar polarizabilities based
on the data of 400 MeV and higher could be affected by uncontrolled model dependencies
and needs to be avoided. Excluding the higher-energy data from the
TAPS analysis could help to resolve the apparent discrepancy between theory and
experiment in \Figref{potato}. 

In \Figref{chiral} we show the pion mass dependence of proton polarizabilities in 
both B$\chi$PT and HB$\chi$PT. The difference between the two for the magnetic polarizability
(lower panel) at $\cO(p^3)$ is stunning (compare the blue dashed  and violet dotted curves).
The region of applicability of the HB expansion is apparently limited here to essentially the chiral
limit, $m_\pi \to 0$. For any finite pion mass, the B$\chi$PT and HB$\chi$PT results
come out to be of a similar magnitude but of the opposite sign. A similar picture is observed
for the $\pi \De$ loops arising at $\cO(p^4/\vDe)$. In fact, we have checked that
in the limit of vanishing $\De$-nucleon mass splitting ($\vDe\to 0$), the considered $\pi N$ and
$\pi \De$ loops give (up to the spin--isospin factors) the same result.

The total effect of the Delta here is the difference between the  $\cO(p^3)$ and the $\cO(p^4/\vDe)$ curves
in B$\chi$PT and the difference between the $\cO(p^3)$ and the $\cO(\eps^3)$ curves in HB$\chi$PT.
Note that here the $\cO(\eps^3)$ contribution
in HB$\chi$PT with $\De$(1232) precisely corresponds to $\cO(p^4/\vDe)$ 
of B$\chi$PT, thus we do not include the $\cO(p^4)$ LECs in neither of the
calculations.  The actual $\cO(\eps^3)$ calculations \cite{Hildebrandt:2003fm} are supplemented with
the $\cO(p^4)$ LECs, whose main  role is then to cancel the large contribution
of the $\De$-isobar.

\section{Results for cross sections}
\seclab{results}

In this section we present the results for the differential cross sections of proton Compton scattering.

\subsection{Unpolarized}

\begin{figure}[bt]
\centerline{ \epsfclipon  
\epsfxsize=10.5cm%
  \epsffile{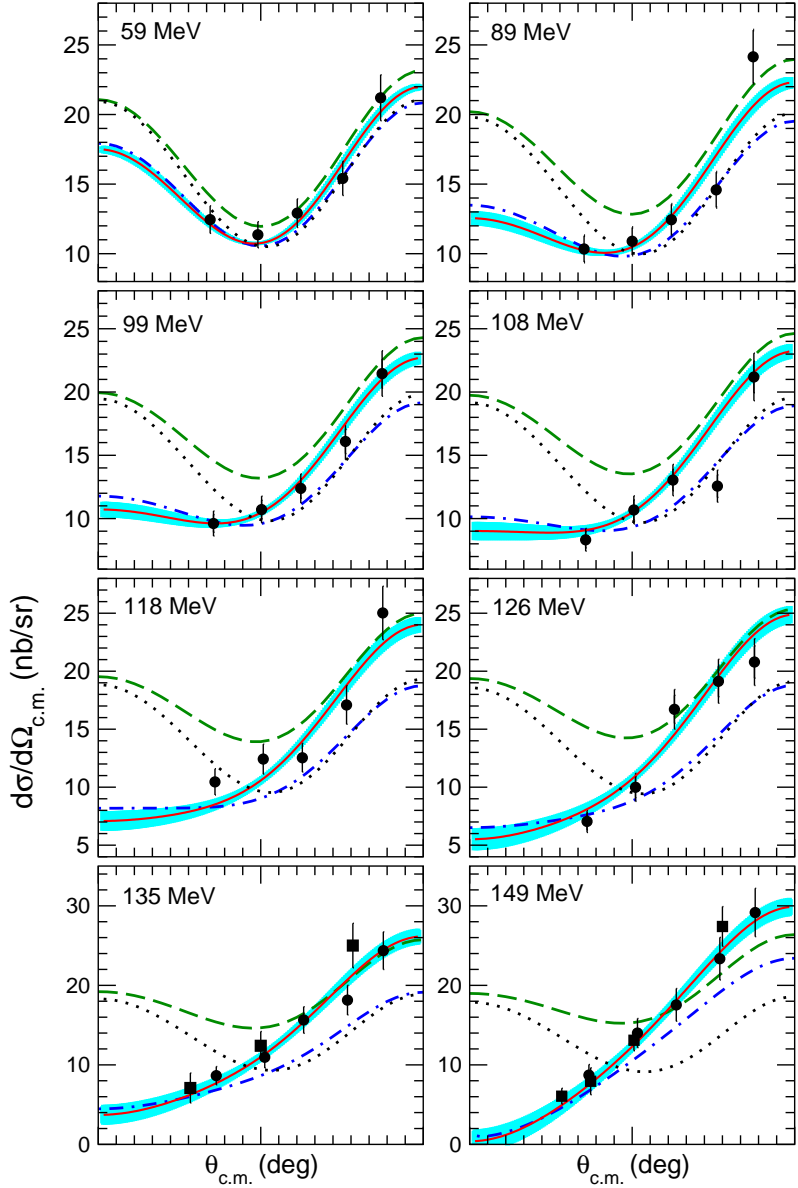} 
}
\caption{(Color online)  Angular dependence of 
the $\ga p\to \ga p$ differential cross section in
the center-of-mass system for a fixed photon-beam energies
as specified for each panel. Data points are from SAL~\cite{Hal93} ---
filled squares, and MAMI~\cite{MAMI01} --- filled circles. The curves are:
Klein--Nishina --- dotted, Born graphs and WZW-anomaly --- green dashed,
adding the $p^3$ $\pi N$ loop contributions of B$\chi$PT
--- blue dash-dotted. The result of adding the $\Delta$
contributions, i.e., the complete NNLO result, is shown by the red solid line with band.}
\figlab{fixE}
\end{figure}

\begin{figure}[bt]
\centerline{\epsfclipon  \epsfxsize=5.6cm%
  \epsffile{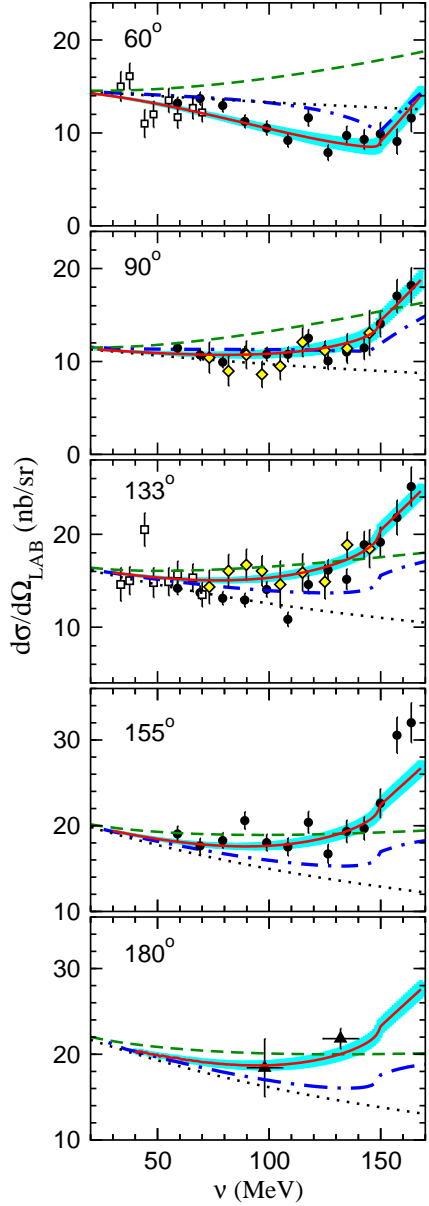} 
}
\caption{(Color online) Energy dependence of 
the $\ga p\to \ga p$ differential cross section in
the lab frame for fixed values of the scattering angle. 
Data points are from: Illinois~\cite{Federspiel:1991yd} --- open squares,
MAMI~\cite{Zieger:1992jq} --- filled triangles, SAL~\cite{MacG95} ---
open diamonds, and MAMI~\cite{MAMI01} --- filled circles. The legend
for the curves is the same as in \Figref{fixE}.} 
\figlab{fixA}
\end{figure}

In \Figref{fixE}, we consider the unpolarized differential cross section
of the $\gamma p\to\gamma p$ process 
as a function of the scattering angle in center-of-mass system, at fixed
incident photon energy. 
 In \Figref{fixA}, we study the same cross section, but as a function
of the energy for fixed scattering angle in the lab frame.
Our complete NNLO result is shown by red solid curves with band indicating
the theory error estimate given in \Eqref{error}. The agreement between the theory and the experiment
is quite remarkable here, especially given the fact that the theory result here is a prediction
in the sense that has no free parameters. Despite this good agreement, as already noted above, 
there is a few-sigma discrepancy in the 
 polarizability values between this theoretical prediction and the most precise
 empirical extraction~\cite{MAMI01}. This is apparently because the data at higher energies
 used additionally in the empirical extraction play an important role
in the determination of $\beta$.

It is always interesting to study the convergence of the chiral expansion.
In these figures the leading-order, $\cO(p^2)$, result is shown by
dotted curve, which is nothing else than the Klein--Nishina cross section
(i.e, Compton scattering off a classical pointlike particle with the charge and mass of the proton). 
The NLO, $\cO(p^3)$, result is given by the blue dash-dotted curve. One can see
that the size of the effects varies strongly with the scattering angle. At energies below 
the pion-production threshold, the NLO effects are tiny at backward angles
but play a crucial role at forward angles. The situation is quite the opposite for the NNLO $\De$-isobar
contributions. Nevertheless, the convergence of this expansion 
seems to be satisfactory and in any case is much better than it would be in analogous HB$\chi$PT
calculations.

For completeness the result for the Born contribution, given by the Powell cross section
together with the WZW anomaly contribution (graphs in \Figref{born}),
is shown here by the green dashed curves. Any deviation from these curves
at low energies is attributed to polarizability effects. More specifically,
\beq
\eqlab{unpolcs}
\frac{d\si}{d\Omega} - \frac{d\si}{d\Omega}^{\!\!(\mathrm{Born})}   = 
- 8 (e^2/4\pi)\,\Phi\,  \nu\nu'     \left[ \half (\alpha+\be) (1+z)^2   -\half(\al- \beta)(1- z)^2  \right] + \cO(\nu^3),
\eeq
where $\Phi$ is defined in \Eqref{phasespace} and $z$ is the cosine of the lab-frame scattering angle.
 
 The difference between
the dashed (Born) curves and the dotted (Klein--Nishina) curves arises mainly due to 
proton's anomalous magnetic moment. The difference is substantial but
at backward angles can be seen to cancel almost entirely against 
the chiral loop contribution at $\cO(p^3)$,
to obtain the blue curves. Thus, at $\cO(p^3)$, there is an intricate cancellation between
the anomalous magnetic moment
 and the chiral loop effects. It would be interesting to see if this cancellation
persists  at higher orders.

\subsection{Polarized}

\begin{figure}[bt]
\centerline{\epsfclipon  \epsfxsize=13cm%
  \epsffile{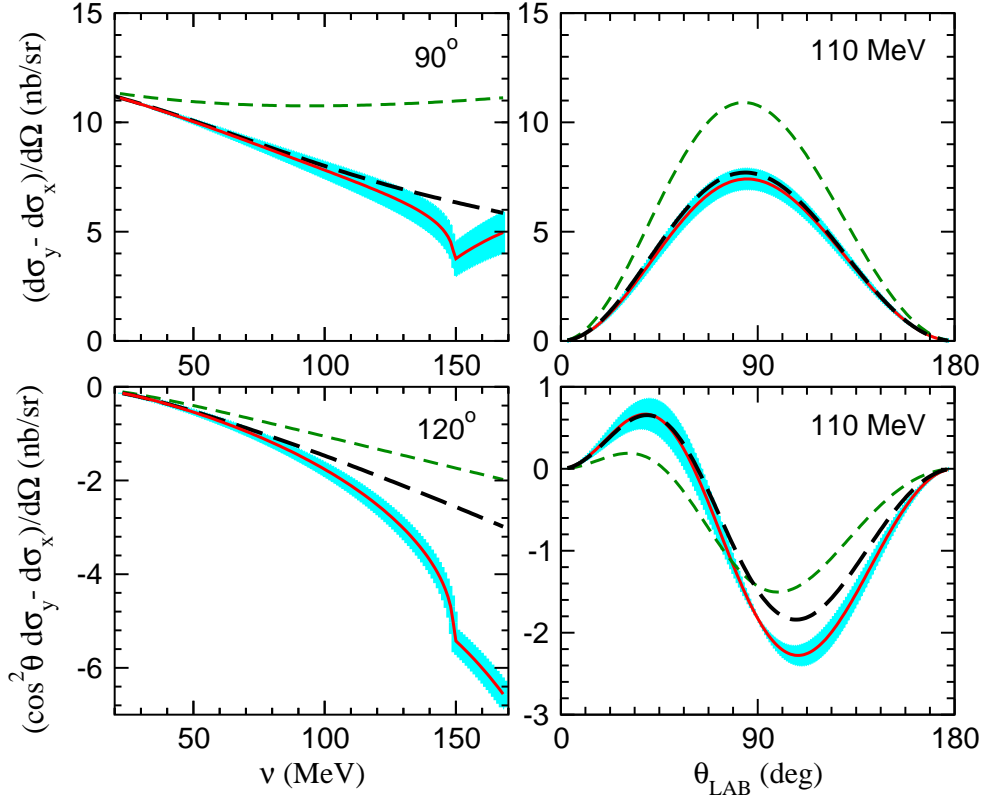} 
}
\caption{(Color online) Linearly polarized
differential cross sections as defined in the text
for fixed values of the scattering angle (left panel) or fixed
energy (right panel).  The green dashed curves ---  Born graphs and WZW-anomaly;
black long-dashed --- in upper panels result of adding $\alpha=10.8$,  
in lower panels result of adding $\beta=4$, both using LEX.
The NNLO result is shown by the red solid line with band.} 
\figlab{pols}
\end{figure}

In \Figref{pols} we show the results for Compton-scattering differential cross sections obtained with linearly polarized beam. The subscript $x$ ($y$) indicates that the beam polarization is parallel (perpendicular)
to the scattering plane, $\th$ is the scattering angle in the lab frame and $d\Omega = -2\pi \sin\th \,d\th $.
The two particular combinations of polarized cross sections, seen in the upper and the lower panels,
are chosen such that each of them is sensitive to only one of the polarizabilities. They
are therefore selected for the forthcoming measurement of proton polarizabilities at the HIGS
facility~\cite{Weller:2009zz}. 
The HIGS measurements are planned to be taken at 110 MeV photon lab energy, where
a low-energy expansion (LEX) is assumed to hold. Indeed a study of the unpolarized differential
cross section indicates that LEX can be trusted to energies of about 100 MeV \cite{MacG95}.
Here we make a similar study for the polarized cross sections.

The LEX states that at the second
order the deviation
of the polarized cross sections from the corresponding Born result (dashed
lines in the figure) is simply given in terms of the polarizabilities:
\begin{subequations}
\eqlab{polcsLEX}
\bea
\frac{d\si_{x}}{d\Omega} - \frac{d\si_{x} }{d\Omega}^{\!\!(\mathrm{Born})} & = & - 8 (e^2/4\pi)\,\Phi\,  \nu\nu' 
\left(\alpha \cos\th  + \beta   \right)\cos\th + \cO(\nu^3) \\
\frac{d\si_{y}}{d\Omega} - \frac{d\si_{y} }{d\Omega}^{\!\!(\mathrm{Born})}  & = &
- 8 (e^2/4\pi)\,\Phi\,  \nu\nu'     \left(\alpha   + \beta \cos\th  \right) + \cO(\nu^3), 
\eea
\end{subequations}
where $\Phi$ is defined in \Eqref{phasespace}. A derivation of these expressions is given
in Appendix B.

From \Eqref{polcsLEX} one can indeed see that, at this order in LEX, the difference of
the polarized cross sections, $d\si_{y}/d\Omega - d\si_{x}/d\Omega$, is proportional to $\alpha$, while the combination $\cos^2\th\, d\si_{y}/d\Omega - d\si_{x}/d\Omega $
is proportional to $\beta$. One should realize, though, that this is only an approximate result which breaks down at sufficiently high energies.
The \Figref{pols} attempts to address this issue in a quantitative way
 by comparing 
 the second-order LEX (long-dashed curves) with the result of NNLO B$\chi$PT (red solid curves with
the error band). The LEX and B$\chi$PT results have exactly the same values for the polarizabilities
$\al$ and $\be$, but the validity of B$\chi$PT extends over the whole considered energy range.

We conclude that a determination  at 110 MeV based on a second-order LEX
can be reliable for $\alpha$, see the upper panels. 
The situation is not as fortunate for the observable aimed at the determination of $\beta$,
see the lower panels. The LEX result begins to fail here at lower energies, 
at least in the backward angles where Compton experiments are usually simpler.

\section{Conclusion}

We have completed a next-to-next-to-leading order (NNLO) 
calculation of low-energy Compton scattering on the proton
within the $\chi$PT framework. More specifically, we have
computed all the effects of order $p^2$, $p^3$, and $p^4/\vDe$,
with $\vDe$ being the excitation energy of the $\De(1232)$ resonance.
These are all {\it predictive} powers in the sense that no unknown low-energy constants (LECs)
enter
until at least one order higher [i.e., $\cO(p^4)$]. This fact together with 
the availability of precise data for Compton scattering has given us a unique
opportunity to put $\chi$PT to a test.

We have found that, assuming a natural size of the $\cO(p^4)$ LECs, the theoretical
uncertainty of the NNLO calculation is comparable with the uncertainty of present
empirical information about the cross sections of proton Compton scattering 
and the corresponding values for isoscalar polarizabilities of the proton.
Within these uncertainties the NNLO result agrees with the cross sections data
below the pion threshold but shows a three-sigma discrepancy in the value for
the magnetic polarizability. We note that the state-of-the-art empirical value for the polarizabilities
was extracted by using not just the low-energy data but also data above the $\De$-resonance 
region. The planned experiments at HIGS could be very helpful in sorting out this issue,
since they plan to use precision low-energy data only. In this case, however,
the reliance on the strict second-order low-energy expansion might be a problem, as
our calculation has shown. The $\chi$PT framework itself could provide a more
reliable energy interpolation needed for the extraction of polarizabilities.

In this work we have insisted on the fact that {\it chiral power counting} should be done
for {\it graphs}, not {\it contributions}. It does not put any constraint
on how many powers of pion mass or energy may appear in the result. 
It puts the constraint on the leading power only.  The heavy-baryon expansion is
therefore not mandatory for correct power-counting. What is important is that
no powers lower than given by power-counting are present in the result.
The manifestly covariant
baryon $\chi$PT (B$\chi$PT) conforms to this requirement, because even if the lower-order
terms appear in calculation of a given graph, they are shown to contribute only
to a renormalization of the LECs. In our example, 
the low-energy theorem and chiral symmetry ensured that all such troublesome
terms contributed only to the renormalization of nucleon mass, charge,
and the anomalous magnetic moment. 

\section*{Acknowledgments} 
We would like to thank Daniel Phillips and Marc Vanderhaeghen for a number of  illuminating discussions,
and Martin Schumacher for a helpful communication. V.~L.\ is grateful to the Institut f{\"u}r
Kernphysik at Johannes Gutenberg Universit{\"a}t Mainz for kind hospitality.

\appendix
\section{Chiral loop contributions to polarizabilities}

Hereby we give the expressions for the loop contributions of $\cO(p^3)$ and $\cO(p^4/\vDe)$ to 
isoscalar proton polarizabilities $\alpha$ and $\beta$, as well as the corresponding
heavy-baryon results.
\subsection{Nucleon Loops}
Our results for the loop contributions at $\cO(p^3)$ agree with
\cite{Bernard:1991rq}:
\begin{eqnarray}
 \alpha&=&\frac{e^2g_A^2}{192\pi^3M_Nf^2}\Bigg\{ -1 + \int\limits^1_0\! \frac{dx}{[D_N(x)]^3}
 \bigg[
2x^4(-3x^3\!+\!8x^2\!-\!9x\!+\!5)\\
&&+\, x^2(9x^4\!-\!26x^3\!+\!29x^2\!-\!18x+7)\mu^2-(9x^5\!-\!33x^4\!+\!45x^3\!-\!27x^2\!+\!7x\!-\!1)\mu^4 \bigg] \Bigg\},\nn \\
\beta &=&\frac{e^2g_A^2}{192\pi^3 M_Nf^2}
\\
& \times & \Bigg\{ 1 - \int\limits^1_0\! \frac{dx}{[D_N(x)]^2} \bigg[ 2x^2(6x^3\!-\!13x^2\!+\!9x\!-\!1) +(9x^4\!-\!24x^3\!+\!21x^2\!-\!6x\!+\!1)\mu^2 \bigg]
\Bigg\}, \nn
\end{eqnarray} 
where $\mu=m_\pi/M_N$,  and $D_N(x)=\mu^2(1-x)+x^2$.

The corresponding heavy-baryon result is obtained from these expressions by expanding
in $\mu$ and keeping the leading term only:
\bea
 \alpha^{(HB)}&=&\frac{10 e^2g_A^2}{768\pi^2 f^2 m_\pi } , \\
\beta^{(HB)}&=& \frac{e^2g_A^2}{768\pi^2 f^2 m_\pi }.
\eea

\subsection{Delta Loops}
The $\cO(p^4/\vDe)$ loops of \Figref{loopsD} give the following contribution to polarizabilities:
\begin{eqnarray}
 \alpha&=&\frac{e^2h_A^2M_N}{3456\pi^3 M_\Delta^2f^2}
\Bigg\{ \frac{25}{2} + 8\de -  3 \int\limits^1_0  \frac{dx\,  {x}^2 }{[D_\Delta(x)]^2}\big[
(1-x) \big(35\!-\!104 x +17 {x^2}\!+\!112 {x^3}\!-\!60 {x^4}\nonumber\\
& & +\, \big(105\!-\!273
x\!+\!72 {x^2}\!+\!92 {x^3}\big) \delta +\big(105\!-\!269 x\!+\!88 {x^2}\!+\!72 {x^3}\big){{\delta }^2}
+\big(35\!-\!100
x\!+\!64 {x^2}\big) {{\delta }^3}\big) \nonumber  \\
&&  - \, x \big(35\!-\!69 x\!-\!40 {x^2}\!+\!72 {x^3}\!+\!\big(35\!-\!100 x\!+\!64 {x^2}\big) \delta \big) {{\mu
}^2}\big]
\nonumber\\
& &-\, 6  \int\limits^1_0  dx\, x  \big(12+9 x-34 {x^2}+3 (4-5 x) \delta \big) \big[\Xi -\log D_\Delta(x)\big]
\Bigg\},\\
\beta &=&\frac{e^2h_A^2M_N}{3456\pi^3  M_\Delta^2f^2}  \Bigg\{ \frac{65}{6} - 8\de
+ \int\limits^1_0  \frac{dx\,  {x}^2 }{D_\Delta(x)}
 (9-32 x+24 {x^2})
(1+x+\delta )\nonumber\\
& &+\,6  \int\limits^1_0  dx\, x  
\big(12+7 x+10 {x^2}+3 (-4+5 x) \delta \big) \big[\Xi -\log D_\Delta(x)\big]
\Bigg\},
\end{eqnarray}
where $\mu=m_\pi/M_N$, $\delta=\vDe/M_N$, and $D_\Delta(x)=(1-x)[(1+\de)^2-x]+x\mu^2$.
Furthermore, in these expressions we have $\Xi=2/(4-d)-\gamma_E+
\log (4\pi\La/M_N)$ the divergence in $d$
dimensions, with $\La$ the dimreg scale. 
Thus, the $\De$ loops contain an ultraviolet divergence  which is to be
renormalized by $\cO(p^4)$ LECs. We choose to define the values for these LECs 
in the $\ol{MS}$ scheme, and hence put $\Xi=0$.

Expanding these results in small $\mu$ and $\delta$ to leading order, 
we reproduce the heavy-baryon result for the $\pi\De$-loop contributions~\cite{Hemmert:1996rw}:
\begin{eqnarray}
 \alpha^{(HB)}&=&\frac{e^2h_A^2}{864\pi^3f^2 \vDe}\left(9+\log\frac{2\vDe}{m_\pi}\right),\\
\beta^{(HB)}&=&\frac{e^2h_A^2}{864\pi^3f^2 \vDe} \log\frac{2\vDe}{m_\pi}\,.
\end{eqnarray}


\section{Low-energy expansion for cross sections}

The differential cross section is given in terms of the Compton amplitude by
\beq
\eqlab{csdef}
\frac{d\si}{dt} = \frac{1}{16\pi (s-M_N^2)^2} \sum_{\la'\si'} \big| T_{\la'\si', \,\la\si}\big|^2\,,
\eeq
where $\la$ and $\si$ are the target and the photon's helicities. The sum is over the final helicities,
the initial ones are fixed.
To find the low-energy expansion (LEX) of this quantity at the second order in energy, we can
ignore the {\it spin-dependent} contribution and write the Compton amplitude as follows:
\beq
T_{\la'\si', \,\la\si} = \left( - A_1(s,t)\, \ceps_{\si'} '\cdot\ceps_\si + A_2(s,t) \, 
q\cdot \ceps_{\si'}'\, q'\cdot\ceps_\si \right) \, 2M_N\,\de_{\la'\la}\,,
\eeq
where $A_i$ are scalar amplitudes dependent on the Mandelstam variables only,
 $q$ ($q'$) is the initial (final) photon 4-momentum, and
\bea
\ceps^\mu & = & \veps^\mu - \frac{P\cdot \veps}{P\cdot q} q^\mu\,, \\
{\ceps'}^{\mu} & = & {\veps'}^{\mu} - \frac{P\cdot \veps'}{P\cdot q'} {q'}^{\mu} 
\eea
with $\veps$ the vectors of photon polarization, and $P=p+p'$ the sum of the nucleon external momenta. Since
\beq
\sum_\si \veps^\mu_\si \veps_\si^{\ast\nu} = -g^{\mu\nu} ,\,\,\, \veps_\si \cdot \veps_\si^{\ast}=-1,
\eeq
we have 
\beq
\sum_\si \ceps^\mu_\si \ceps_\si^{\ast\nu} = -g^{\mu\nu} +\frac{P^\mu q^\nu +P^\nu q^\mu}{P\cdot q}
- \frac{P^2 q^\mu q^\nu}{(P\cdot q)^2}
\eeq
with
$
P\cdot q  = \half(s-M_N^2 - u+M_N^2)=M_N (\nu+\nu')
$.
We therefore obtain
\bea
\sum_{\la'\si'} \big| T_{\la'\si', \,\la\si}\big|^2  & = & (2M_N)^2
(-A_1 \ceps_{\si\mu} + A_2 q_\mu q'\cdot \ceps_\si) (-A_1 \ceps_{\si\nu}^\ast + 
A_2 q_\nu q'\cdot \ceps_\si^\ast ) \nn\\
& \times & \big(  -g^{\mu\nu} +\frac{P^\mu {q'}^\nu +P^\nu {q'}^\mu}{P\cdot q}
- \frac{P^2 {q'}^\mu {q'}^\nu}{(P\cdot q')^2} \big). \nn 
\eea

We next take the Born contribution out of $A_1$:
\beq
\tilde{A}_1 = A_1 +  e^2/M_N
\eeq
and use the LEX:
\bea
\tilde{A}_1 &=& 4\pi  (\alpha + \beta z) \, \nu \nu' + \cO(\nu^3)\,,\\
\tilde A_2 &=& -4 \pi \beta + \cO(\nu)\,,
\eea
  where $z$ is the cosine of the lab-frame scattering angle, $\al$ and $\be$ are respectively the
  electric and the magnetic polarizability.

At the second order in $\nu$ for the non-Born (NB) contribution we thus have
\bea
\sum_{\la'\si'} \big| T^{(NB)}_{\la'\si', \,\la\si}\big|^2 & =&  -8 M_N (4\pi e^2) \nu\nu'
\left[ \alpha + \beta z + (\hat q' \cdot \eps_\si)^2  \beta 
- \frac{4\nu\nu'}{(\nu+\nu')^2}  (\hat q' \cdot \eps_\si)^2 (\al +\be) \right] \nn\\
& =&  -8 M_N (4\pi e^2) \nu\nu'
\left[ \big(1- (\hat q' \cdot \eps_\si)^2\big)\alpha + \beta z \right] +\cO(\nu^3)\,,
\eea
where $\hat{q}\,' = (1,\sqrt{1-z^2}, 0 , z)$. 
Substituting in \Eqref{csdef} and 
selecting the appropriate photon polarization we arrive at  \Eqref{polcsLEX}.

\end{document}